\documentstyle[aps,manuscript,epsfig]{revtex}
\begin{document}
\topmargin 0.cm
\textwidth 15.cm
\textheight 23.cm

\title{Evidence for a Soft Nuclear Equation-of-State\\
from Kaon Production in Heavy Ion Collisions}
 
\author{
C.~Sturm$^a$,
I.~B\"ottcher$^d$,
M.~D\c{e}bowski$^e$,
A.~F\"orster$^a$,
E.~Grosse$^{f,g}$,
P.~Koczo\'n$^b$,
B.~Kohlmeyer$^d$,
F.~Laue$^{b,\dag}$, 
M.~Mang$^b$,
L.~Naumann$^f$,
H.~Oeschler$^a$,
F.~P\"uhlhofer$^d$,
E.~Schwab$^b$,
P.~Senger$^b$,
Y.~Shin$^c$,
J.~Speer$^d$,
H.~Str\"obele$^c$,
G.~Surowka$^{b,e}$,
F.~Uhlig$^a$
A.~Wagner$^{h,f}$,
W.~Walu\'s$^e$\\
(KaoS Collaboration)\\ 
$^a$ Technische Universit\"at Darmstadt, D-64289 Darmstadt, Germany\\
$^b$ Gesellschaft f\"ur Schwerionenforschung, D-64220 Darmstadt, Germany\\
$^c$ Johann Wolfgang Goethe Universit\"at, D-60325 Frankfurt am Main, Germany\\
$^d$ Phillips Universit\"at, D-35037  Marburg, Germany\\
$^e$ Uniwersytet Jagiello\'{n}ski, PL-30-059\,Krak\'ow, Poland\\
$^f$ Forschungszentrum Rossendorf, D-01314 Dresden, Germany \\
$^g$ Technische Universit\"at Dresden, D-01314 Dresden, Germany\\
$^h$ National Superconducting Laboratory, Michigan State University, East
Lansing, Michigan 48824, USA  \\
}
\maketitle


\vspace{-1.cm}
\begin{abstract}
The production of pions and kaons has been measured in $^{197}$Au+$^{197}$Au
collisions at beam energies from 0.6 to 1.5 A$\cdot$GeV with the 
Kaon Spectrometer at SIS/GSI. The $K^+$ meson multiplicity per  
nucleon is enhanced in Au+Au collisions by factors up to 6 relative to 
C+C reactions whereas the corresponding $\pi$ ratio is reduced. 
The ratio of the $K^+$ meson excitation functions for Au+Au and
C+C collisions increases with decreasing beam energy. 
This behavior is expected for a soft nuclear equation-of-state.

\end{abstract}

\newpage

The equation-of-state of nuclear matter plays an important role in
the dynamics of a supernova explosion and the stability of neutron stars.
So far our knowledge on the behavior of dense nuclear matter
is based on the study of nuclear excitations \cite{young}
and an extrapolation to high densities.   
Nucleus-nucleus collisions at relativistic energies offer the unique 
possibility to study experimentally the properties of dense nuclear matter 
in the laboratory. The collective flow of nucleons \cite{reis_ritt,daniel}
and the yield of produced mesons  \cite{stock} were considered to be
sensitive to the compressibility of nuclear matter. 
For example, in pioneering experiments at the BEVALAC the   
production of pions was studied in heavy ion collisions,  and the pion  
multiplicity was  correlated to the  thermal energy of the fireball
in order to extract information on the equation-of-state of nuclear matter 
\cite{harris}.

Microscopic transport calculations indicate that the yield  of kaons created
in collisions between heavy nuclei at subthreshold beam energies 
(E$_{beam}$ = 1.58 GeV for NN$\to$$K^+\Lambda$N)
is sensitive to  the compressibility of
nuclear matter at high baryon densities \cite{aich_ko,li_ko}. 
This sensitivity is due to the production mechanism of K$^+$ mesons. 
At subthreshold  beam energies,
the production of kaons requires multiple nucleon-nucleon collisions 
or secondary collisions  such as $\pi$N$\to$$K^+\Lambda$. These processes
are expected to occur predominantly at high baryon densities, and
the densities reached in the fireball depend on the nuclear equation-of-state
\cite{fuchs97}.

$K^+$ mesons are well suited to probe the properties of the 
dense nuclear medium because of their long mean free path. 
The propagation of  $K^+$ mesons in nuclear matter is characterized 
by the absence of  absorption  
(as they contain an antistrange quark) and hence kaons emerge as 
messengers from  the dense phase of the collision. 
In contrast, the pions created in the high density phase of the 
collision are likely to be reabsorbed and most of them  
will leave  the reaction zone in the late phase 
\cite{bass,wagner1}.

The influence of the medium on the $K^+$ yield is amplified 
by the steep excitation function of kaon production near threshold energies.
Early transport calculations find that the $K^+$ yield from Au+Au 
collisions at subthreshold energies will be  enhanced by a factor of about  
2 if a soft rather than a hard equation-of-state is assumed
\cite{aich_ko,li_ko}. Recent calculations take into account the modification
of the  kaon properties in the dense nuclear medium \cite{ko_li,cass_brat}.
When  assuming a repulsive $K^+$N potential as proposed by various 
theoretical models (see \cite{schaffner} and references therein) 
the energy needed to create a $K^+$ meson in the  nuclear medium is 
increased  and hence the $K^+$ yield will be reduced. 
Therefore, the yield of $K^+$ mesons  produced in heavy ion collisions
is affected by both the nuclear compressibility and 
the in-medium kaon potential. 

Our idea is to disentangle these  two competing effects by studying 
$K^+$ production in a very light ($^{12}$C+$^{12}$C) 
and a heavy collision system ($^{197}$Au+$^{197}$Au) 
at different beam energies near threshold. 
The reaction volume is more than 15 times  larger  in Au+Au  
than in  C+C collisions and hence the average baryonic density - achieved 
by the pile-up of nucleons - is significantly higher \cite{cass_brat}.  
Moreover, the maximum baryonic density reached in Au+Au collisions  
depends on the nuclear compressibility \cite{li_ko,aichelin} 
whereas in the small C+C system this dependence is very weak \cite{fuchs00}.
The repulsive $K^+$N potential is assumed to depend nearly
(or less than) linearly on the baryonic density \cite{schaffner} and thus
reduces the kaon yield accordingly. 
On the other hand, at subthreshold beam energies 
the $K$ mesons are created in secondary collisions involving two or more 
particles and hence the production of $K^+$ mesons depends 
at least quadratically on the density.
These multiple-step kaon production processes contribute increasingly 
with decreasing beam energy. Therefore, the $K^+$ production 
excitation function in Au+Au collisions is expected to 
be influenced stronger by the nuclear compressibility than by the in-medium
potential. In contrast,
the $K^+$ production processes in C+C collisions are expected to
be little  affected by the nuclear equation-of-state.
Therefore, the comparison of precision data on $K^+$ production  
as function of beam energy in C+C and Au+Au collisions should reveal 
effects caused by the compressibility of nuclear matter rather than 
in-medium modifications of the $K^+$ mesons.     

Data on $K^+$ production in heavy ion collisions at beam energies around 
and below the nucleon-nucleon threshold are still scarce.  
The creation of $K^+$ mesons  in Au+Au collisions at 
1 A$\cdot$GeV has been investigated in an early experiment with 
large statistical errors \cite{misko}. Production 
cross sections of  $K^+$ mesons have been measured in Ni+Ni collisions at 
several beam energies \cite{barth,best}. Data on  $K^+$  
production in C+C collisions at beam energies from 0.8
to 2.0 A$\cdot$GeV have been published recently \cite{laue}.        

In this Letter we report on a series of experiments which were performed 
in order to study systematically the production of pions and kaons 
as function of the beam energy in   symmetric nucleus-nucleus collisions 
using  light and very heavy nuclei.  We measured differential production
cross sections for $K^+$ and $\pi^+$ mesons in Au+Au collisions from
0.6 to 1.5 A$\cdot$GeV at different polar emission angles. 
The  experimental results are compared to inclusive cross sections for 
kaon and pion production in C+C collisions (the pion data are shown in
this Letter the first time).

The experiments were performed with the Kaon Spectrometer (KaoS) at the
heavy ion synchrotron (SIS) at GSI in Darmstadt \cite{senger}.
The magnetic spectrometer has a large acceptance
in solid angle and momentum ($\Omega\approx$30 msr, $p_{max}/p_{min}\approx$2).
The short distance of 5 - 6.5 m from target
to focal plane minimizes kaon decays in flight.
Particle identification and the trigger are based on separate measurements of
velocity, momentum and time-of-flight. The trigger suppresses pions and
protons by factors of 10$^2$ and 10$^3$, respectively.
The background due to spurious tracks and pile-up is
removed by trajectory reconstruction based on
three large-area multi-wire proportional counters. 
The  remaining background below
the kaon mass peak ($\approx$ 10\% in Au+Au at 1.46 A$\cdot$GeV and 
$\approx$ 30\% at 0.78 A$\cdot$GeV) is subtracted. The loss of kaons
decaying in flight is determined and accounted for
by Monte Carlo simulations using the GEANT code.

The $^{197}$Au beam had  an intensity of about 5$\times10^7$ ions per spill.
The $K$ mesons were registered 
at polar angles between  $\Theta_{lab}$ = 40$^{\circ}$ and 84$^{\circ}$
over a momentum range of 260 $<$ $p_{lab}<$ 1100 MeV/c.
The (approximate) raw numbers of detected $K$ mesons are listed in Table 1.

Figure 1 shows the inclusive production cross section for $K^+$ mesons
as function of the laboratory momentum
measured in Au+Au collisions at different beam energies and 
laboratory emission angles.
After correction for the energy loss in the Au target 
(thickness 0.5 and 1.0 mm) the average beam energies are reduced 
to 0.56, 0.78, 0.96, 1.1 and 1.46 A$\cdot$GeV. 
The error bars shown are due to statistical uncertainties 
and background subtraction. An overall systematic error of 10 \%
due to efficiency corrections and beam normalization 
has to be added.
The solid lines represent the function 
\begin{equation}
d^3\sigma/dp^3 = C \cdot (1+a_2 \cdot cos^2\Theta)\cdot exp(-E/T)
\end{equation}
which is  fitted to the data in the center-of-mass (c.m.) system   
and transformed into the laboratory. $C$ is a normalization constant,
$a_2$ parameterizes the anisotropy of the polar angle distribution in 
the c.m. system and the exponential describes the energy distribution 
(with $T$ the inverse slope parameter). The function  is fitted simultaneously
to the spectra measured  at different angles at a given beam energy. 
The parameters $a_2$ are determined by the fit
for the beam energies of 0.78, 0.96 and 1.46 A$\cdot$GeV;
at 0.56 and 1.1 A$\cdot$GeV the $a_2$ values are 
estimated by extra- and interpolation, respectively.   
The resulting values for $T$ and $a_2$ are listed in Table 1. 
The polar angle distribution is found to be 
forward-backward peaked in the c.m. system, and the values of $a_2$
correspond to a nonisotropic contribution of less than 30\% 
(in C+C collisions the  nonisotropic fraction is 20\% \cite{laue}). 
The effect of the parameter $a_2$ on the laboratory spectra 
is demonstrated for the beam energy of 0.96 A$\cdot$GeV by the dashed lines in 
Figure 1. These lines represent calculations using equation (1) with
$a_2$=0, i.e. with an isotropic angular distribution. The isotropic 
extrapolation  underestimates the kaon data taken at 
$\Theta_{lab}$=84$^{\circ}$ by a factor of about 2.           

Inclusive  kaon production cross sections are determined by
extrapolations to the non measured phase-space regions using equation (1) 
with the parameters as obtained by the fits to the spectra. 
Particle multiplicities are calculated using the 
inclusive $\pi^+$ and $K^+$ production cross sections according to
$M = \sigma/\sigma_R$ with  $\sigma_R$ the geometrical 
cross section of the reaction 
($\sigma_R$ = 4$\pi$(1.2 A$^{1/3}$)$^2$ fm$^2$ = 0.95 b for C+C and 6.1 b
for Au+Au collisions). The latter value is in agreement with a minimum
bias measurement \cite{wagner2}. 

Figure 2 (upper panel)  shows the pion and $K^+$ multiplicity per 
nucleon for C+C and Au+Au collisions as a function of beam energy.
The error bars include systematic uncertainties due to the extrapolation to 
full phase space and due to beam normalization and efficiencies.  
The pion data points are scaled by a factor of 1/100; they represent the 
sum of charged and neutral pions as 
calculated from the measured $\pi^+$ multiplicities
according to the isobar model \cite{arndt}. This model $-$ which 
explains very well the $\pi^+/\pi^-$ ratios measured in Au+Au collisions 
\cite{wagner3} $-$ gives a 
ratio of all pions to positively charged pions    
of $\pi^{all}/\pi^+$ = 4.4 (at 0.56 A$\cdot$GeV) and 
4.1 (at 1.46 A$\cdot$GeV) for Au+Au and  $\pi^{all}/\pi^+$ = 3 
for C+C collisions.  
The pion multiplicity per nucleon is smaller in Au+Au than in 
C+C collisions whereas the $K^+$ multiplicity per nucleon
is larger. This observation demonstrates that the nuclear medium 
- as formed in the heavy system - 
affects the production (and absorption) of pions and kaons 
in a very different manner.

In order to illustrate the different behaviour of pions and kaons 
in nuclear matter we plot the ratio of the pion and kaon excitation functions 
$(M/A)_{Au+Au}/(M/A)_{C+C}$ for Au+Au and C+C in the 
lower panel of Figure 2.  
Due to the different energy losses of the Au and C projectiles in the 
respective targets, the excitation functions for Au+Au and C+C collisions
are not measured exactly at the same effective beam energies. The ratio
$(M/A)_{Au+Au}/(M/A)_{C+C}$ is determined at 0.8, 1.0, 1.2 and 1.46 
A$\cdot$GeV. At 0.8 - 1.2 A$\cdot$GeV we take the meson data measured 
in C+C collisions and interpolate the results from  Au+Au collisions. 
At 1.46 A$\cdot$GeV we take the Au+Au data and interpolate the C+C data. 
For the interpolation we use the fit functions shown as lines  
in the upper panel of Figure 2. The error bars of the kaon multiplicity ratios
include systematic uncertainties due to the extrapolation procedure. The
experimental uncertainties due to efficiencies, acceptances and beam 
normalization, however, cancel in the ratio and therefore have not been 
taken into account.

The pion ratio $(M/A)_{Au+Au}/(M/A)_{C+C}$ (full triangles) 
is smaller than unity and 
decreases by a factor of about 1.7 with decreasing 
beam energy (from 1.46 to 0.8 A$\cdot$GeV).
A ratio smaller than unity might be caused by the reabsorption  
of pions which is more effective in the larger system. 
In principle, the decompressional flow of nuclear matter - 
which is expected to be more important in Au+Au than in C+C collisions - 
reduces the energy available for particle production and therefore may 
contribute to the pion deficit as well.  
This argument is also valid for the kaons; in this case, however, the opposite 
effect is observed.  
   
In contrast to the pion data, the kaon ratio $(M/A)_{Au+Au}/(M/A)_{C+C}$ 
increases by a factor of almost 3  with decreasing  beam energy.
An increase of the $K^+$ yield with decreasing beam energy  
is found by a transport model calculation in central Au+Au collisions if a soft 
instead of  a hard  equation-of-state is used \cite{li_ko}. 
The sensitivity of the kaon multiplicity on the nuclear 
compressibility is enhanced at beam energies well below
the kaon production threshold  because 
the  energy required to create a $K^+$ meson has to be accumulated by
multiple collisions of the participating nucleons.

In order to exclude trivial phase-space effects as the reason for the 
observed effect we present in the lower panel of Figure 2  
the ratio $(M/A)_{Au+Au}/(M/A)_{C+C}$
for pions with kinetic c.m. energies above E$_{kin}^{cm}$ = 0.6 GeV. 
The production of these pions is equivalent $-$ in terms of available 
energy $-$ to the production of $K^+$ mesons with a kinetic energy
of 70 MeV. At this energy the kaon yields have reached their 
maximum values.

Both pions and kaons are created with three particles in the final state
(NN$\pi$ and $K\Lambda$N, respectively), and thus we consider the yield of 
high-energy pions to reflect the phase-space available for particle production. 
Figure 2 (lower panel) 
demonstrates that the  high-energy pion data (open squares)
do not show any beam-energy dependence. Therefore, the increase of the 
kaon ratio (full circles)  with decreasing beam energy is not caused by  
phase-space effects.

Recent QMD transport 
calculations which  take into account a repulsive kaon-nucleon potential
reproduce the energy dependence of the kaon ratio  
as presented in Figure 2 if a compression modulus  of $\kappa$ = 200 MeV 
for nuclear matter is assumed \cite{fuchs00}. 
These calculations use momentum-dependent Skyrme forces to determine
the compressional energy per nucleon (i.e. the energy stored in compression)
as function of nuclear density.  
For a compression modulus  of  $\kappa$ = 380 MeV 
(a ''hard'' equation-of-state)  the calculations find a
kaon ratio $(M/A)_{Au+Au}/(M/A)_{C+C}$ which is below a value of 2 in the 
beam energy range considered \cite{fuchs00}. 

In summary, we have compared pion and kaon production cross sections measured
in C+C collisions at beam energies from 0.8 to 2.0 A$\cdot$GeV
and in Au+Au collisions from 0.6 to 1.5 A$\cdot$GeV. 
At the beam energy of 0.8 A$\cdot$GeV, the $K^+$ meson multiplicity
per nucleon is about a  factor of 6 larger  in Au+Au 
than in C+C collisions  whereas the pion 
multiplicity per nucleon is smaller by a factor of about 2.
The multiplicity of high-energy pions per nucleon
is nearly independent of collision system size and beam energy. 
These high-energy pions  are considered to be a 
reference for the phase-space available for kaon production. 
Therefore, the observed enhancement of the kaon yield per nucleon 
in Au+Au collisions as compared to C+C collisions is not due to a smaller 
phase-space volume in the light system but rather due to collective
effects which do not affect pion production in the same way.
Our data indicate that the  baryonic densities reached in the heavy system 
are as high as expected for a soft nuclear equation-of-state.

The authors acknowledge fruitful discussions with J. Aichelin and C. Fuchs. 
This work is supported by the Bundesministerium f\"ur Bildung und Wissenschaft,
Forschung und Technologie under contract 06DA819 and by the
Gesellschaft f\"ur Schwer\-ionen\-forschung under contract DA\,OESK and 
MR\,PUK and by the Polish Commitee of Scientific Research under contract No.
2P3B11515.

$\dag$ Present address: Ohio State University, Columbus, OH-43210, USA

\newpage
\begin{table}
Table 1: Sample sizes N, inverse slope parameters $T$, 
anisotropy parameters $a_2$ 
and inclusive production cross sections $\sigma$ 
for $K^+$ mesons in Au+Au collisions. 
The values for $T$, $a_2$ and $\sigma$ are determined
by fitting the function defined in equation (1) to the data (see text). 
\begin{center}
\begin{tabular}{|c|c|c|c|c|c|}
E$_{beam}$ (A$\cdot$GeV)  & $\Theta_{lab}$ &N &$T$ (MeV)&$a_2$ & $\sigma$ (mb)\\
\hline
0.56  & 50$^{\circ}$   & 300 & 49$\pm$8& 1.06$\pm0.5$ & 0.5$\pm$0.1 \\
0.78 & 44$^{\circ}$, 84$^{\circ}$ & 1200& 67$\pm$4 & 1.08$\pm0.38$& 8.5$\pm$1.2\\
0.96 & 44$^{\circ}$, 84$^{\circ}$  & 3200 & 82$\pm$4 &1.05$\pm0.22$& 31$\pm$4.0\\
1.1  & 56$^{\circ}$ &1100 & 90$\pm$6 & 1.06$\pm0.4$& 64$\pm$10 \\
1.46  & 40$^{\circ}$, 48$^{\circ}$, 56$^{\circ}$ &2900
& 100$\pm$5 & 1.06$\pm0.3$& 267$\pm$30 \\
\end{tabular}
\end{center}
\end{table}

\newpage
Figure 1:

Inclusive $K^+$ production cross-section
as function of the laboratory momentum  measured in Au+Au collisions
at beam energies of 0.78, 0.96, 1.10 and 1.46 A$\cdot$GeV at various 
laboratory angles. The solid lines represent fits to the data   
according to equation (1) assuming a nonisotropic polar emission pattern 
and the dashed lines at 0.96 A$\cdot$GeV illustrate 
calculations assuming isotropic kaon emission (see text).

\vspace{2.cm}
Figure 2:
Upper panel:
Pion and $K^+$ multiplicity per nucleon $M/A$ 
for Au+Au and C+C collisions as function of the projectile energy per nucleon.  
The pion multiplicities include all pions (see text). The lines 
represent a fit to the data.  

Lower panel:
Ratio of the multiplicities per nucleon  
(Au+Au over C+C collisions)
for $K^+$ mesons (full circles), pions (full triangles) and high-energy
pions (E$^{cm}_{kin}>$ 0.6 GeV, open squares) as function of the projectile 
energy per nucleon.

\begin{figure}[h]
\mbox{\epsfig{file=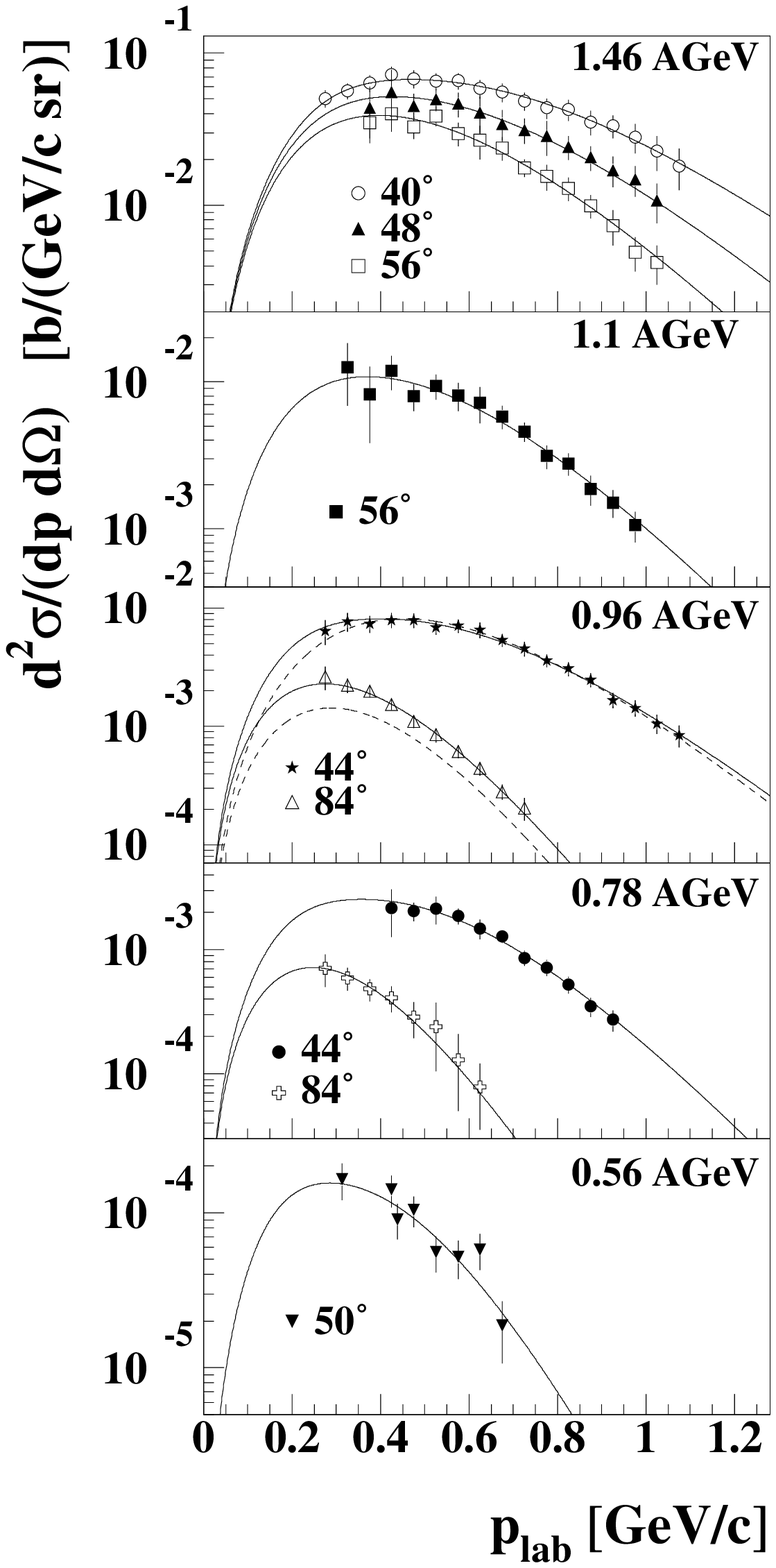,width=12.cm}}
\caption{}
\end{figure}

\begin{figure}[h]
\mbox{\epsfig{file=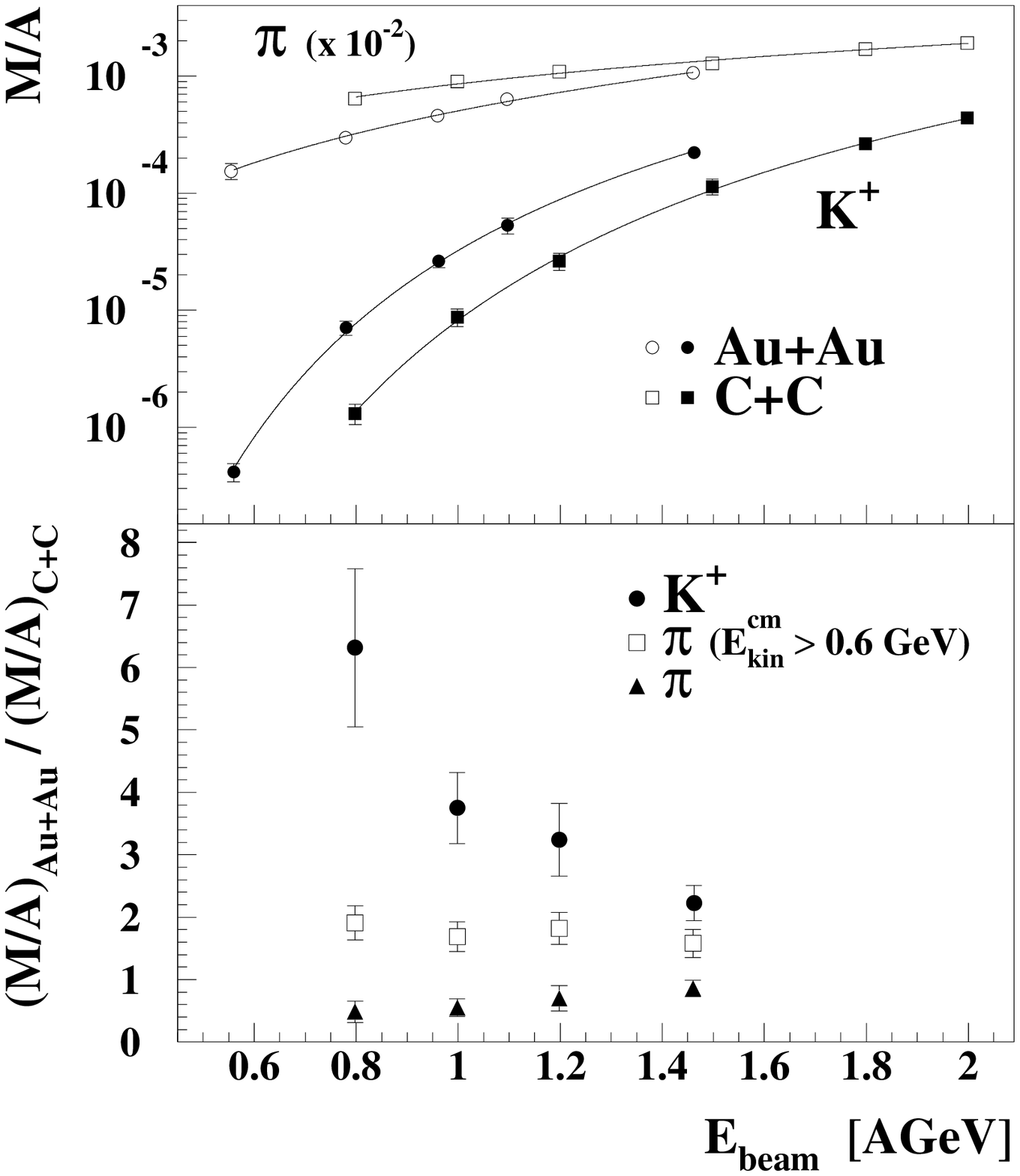,width=16.cm}}
\caption{}
\end{figure}                

\end{document}